\documentclass[conference]{IEEEtran}
\IEEEoverridecommandlockouts
\usepackage{cite}
\usepackage{algorithmic}
\usepackage{textcomp}
\usepackage{xcolor}
\usepackage{soul}
\usepackage{graphicx,color}
\usepackage{amsfonts,amsthm,amssymb}
\usepackage{mathrsfs,amsmath,arydshln,latexsym}
\usepackage{cite,url}
\usepackage[bookmarks,colorlinks]{hyperref}
\usepackage[linesnumbered,ruled,vlined]{algorithm2e}
\usepackage{multirow,array,makecell}
\usepackage{stfloats}
\usepackage{balance}
\usepackage{fancyhdr}
\usepackage{bm}
\usepackage{enumitem}
\usepackage{pifont}
\usepackage{subcaption}
\usepackage{nomencl}
\usepackage{colortbl}
\usepackage{diagbox}
\usepackage{balance}

\allowdisplaybreaks[4]

\begin{document}

\title{A Transmit-Receive Parameter Separable Electromagnetic Channel Model for\\ LoS Holographic MIMO
	\vspace{-0.3em}
}

\author{Tierui Gong, 
	Chongwen Huang,
	Jiguang He,
	Marco Di Renzo,
	Mérouane Debbah,
	and Chau Yuen 
	\vspace{-1em}
	\thanks{T. Gong and C. Yuen are with the School of Electrical and Electronics Engineering, Nanyang Technological University, Singapore (e-mails: trgTerry1113@gmail.com, chau.yuen@ntu.edu.sg).
	C. Huang is with the College of Information Science and Electronic Engineering, Zhejiang University, Hangzhou 310027, China (e-mails: chongwenhuang@zju.edu.cn). 
	J. He is with the Technology Innovation Institute, Masdar City 9639, Abu Dhabi, United Arab Emirates (e-mails: jiguang.he@tii.ae). 
	M. D. Renzo is with the Université Paris-Saclay, CNRS, CentraleSupélec, Laboratoire des Signaux et Systèmes, 3 Rue Joliot-Curie, 91192 Gif-sur-Yvette, France (e-mails: marco.di-renzo@universite-paris-saclay.fr). 
	M. Debbah is with Khalifa University of Science and Technology, Abu Dhabi 127788, United Arab Emirates, and also with the CentraleSupelec, University ParisSaclay, 91192 Gif-sur-Yvette, France (e-mail: merouane.debbah@ku.ac.ae).}
 \thanks{This work was supported in part by the Ministry of Education, Singapore, under its MOE Tier 2 (Award number MOE-T2EP50220-0019), and in part by the Science and Engineering Research Council of A*STAR (Agency for Science, Technology and Research) Singapore, under Grant No. M22L1b0110. 
 The work of Chongwen Huang was supported in part by the China National Key R\&D Program under Grant 2021YFA1000500, in part by the National Natural Science Foundation of China under Grants 62331023 and 62101492, and in part by the Zhejiang Provincial Natural Science Foundation of China under Grant LR22F010002.
 The work of Marco Di Renzo was supported in part by the European Commission through the H2020 ARIADNE project under grant agreement number 871464 and through the H2020 RISE-6G project under grant agreement number 101017011, and by the Agence Nationale de la Recherche (ANR PEPR-5G and Future Networks, NF-SYSTERA 22-PEFT-0006).}
}

\maketitle

\begin{abstract}
	To support the extremely high spectral efficiency and energy efficiency requirements, and emerging applications of future wireless communications, holographic multiple-input multiple-output (H-MIMO) technology is envisioned as one of the most promising enablers. It can potentially bring extra degrees-of-freedom for communications and signal processing, including spatial multiplexing in line-of-sight (LoS) channels and electromagnetic (EM) field processing performed using specialized devices, to attain the fundamental limits of wireless communications. In this context, EM-domain channel modeling is critical to harvest the benefits offered by H-MIMO. Existing EM-domain channel models are built based on the tensor Green function, which require prior knowledge of the global position and/or the relative distances and directions of the transmit/receive antenna elements. Such knowledge may be difficult to acquire in real-world applications due to extensive measurements needed for obtaining this data. To overcome this limitation, we propose a transmit-receive parameter separable channel model methodology in which the EM-domain (or holographic) channel can be simply acquired from the distance/direction measured between the center-points between the transmit and receive surfaces, and the local positions between the transmit and receive elements, thus avoiding extensive global parameter measurements. Analysis and numerical results showcase the effectiveness of the proposed channel modeling approach in approximating the H-MIMO channel, and achieving the theoretical channel capacity.
\end{abstract}


\section{Introduction}

Holographic multiple-input multiple-output (H-MIMO) is considered as one of the most promising directions to enable future sixth-generation (6G) wireless communications. H-MIMO uncovers the possibility of performing advanced signal processing operations in the electromagnetic- (EM-) domain by using specialized devices, and paves the way to realize holographic imaging-level communications \cite{Gong2023Holographic}. 
The raising interest in H-MIMO is essentially due to the extreme performance requirements of 6G, and the recent developments in advanced antenna technologies and new materials.  

Specifically, the extreme requirements of 6G in terms of spectral efficiency and energy efficiency \cite{Saad2020Vision} impose the need for developing wireless transmission schemes from a more fundamental EM perspective, in order to attain the fundamental physical limits of communications.  The extension of the mathematical notions of information/communication theory and statistical signal processing to incorporate the notion of physics of wave propagation, which is referred to as electromagnetic signal and information theory (ESIT), has recently emerged as a possible framework for H-MIMO. ESIT is a broad research field that is concerned with the mathematical treatment and information processing of electromagnetic fields governing the transmission and processing of messages through communication systems. ESIT is more physically-consistent than a mathematical information theoretic framework that treats the wireless channel as a conditional probability distribution, neglecting its inherent physical effects. 

Recent advances in antenna technologies and materials are another important driving factor of H-MIMO. Newly emerging antennas are implemented using advanced metamaterials, metasurfaces, and sub-wavelength and electromagnetically coupled antennas, which offer unprecedented capabilities to manipulate the EM waves \cite{Huang2020Holographic}. 
The antenna apertures are envisioned to be electromagnetically larger than in the past, and to be equipped with a densely deployed number of radiating elements, which enable advanced EM manipulations, including the surface waves. These emerging antenna designs present additional advantages, such as low-cost and low power consumption compared to conventional massive MIMO hybrid architectures \cite{Gong2020RF}, which potentially results in increasing the energy efficiency of future communications \cite{Zappone2022Surface}. These implementation advantages facilitate the fabrication of electrically-large apertures, which enable extremely large beamforming gains, help overcoming the high path loss at high frequency bands, and shift the regime of wireless transmission from the far field to the middle field, or even the near field, facilitating spatial multiplexing in low scattering environments \cite{Di2022MIMO,Bartoli2023Spatial}.

To exploit the potential benefits of H-MIMO, a critical aspect is accurately modeling the EM-domain channel. In wireless communications, the wireless channel is usually modeled through mathematical abstractions, which include the Rayleigh fading model \cite{Tse2005Fundamentals}, the correlated Rayleigh fading model \cite{Bjornson2018Massive}, and the cluster-based geometric model \cite{Heath2016Overview}. These models are usually applied when the transmitter (TX) and receiver (RX) are in the far field of each other. The electrically-large aperture of H-MIMO enables middle field and/or near field communications, which require different models that account for the spherical wavefront of the EM waves as opposed to the plane wave wavefront of far-field channel models typically employed in wireless communications \cite{Lu2022Communicating, Zhang2022Beam}. In this context, purely mathematical-based channel models, even though useful, may be insufficient to completely capture the essence and the opportunities offered by H-MIMO.

EM-native channel models for H-MIMO are necessary, and have recently been proposed. In \cite{Pizzo2020Spatially} and \cite{Pizzo2022Fourier}, the authors considered the channel responses as a spatially-stationary EM random field, which can be statistically depicted by a four-dimensional Fourier plane-wave representation. This channel model is suitable for modeling the far field small-scale fading. Aiming at devising a computationally-efficient line-of-sight (LoS) channel model departing from its integral definition, the authors of \cite{Gong2023Generalized,Gong2023HMIMO} proposed a generalized EM-domain channel model for H-MIMO that can be applied to arbitrary surface placements. This channel model requires the global position information of the antenna elements and/or all relative distances and directions of the TX-RX antenna pairs, which is, however, often inconvenient in practice, especially for a large number of antenna elements. This is because one needs to measure all the global positions and/or all relative distances to obtain the channel.

In this article, we introduce an efficient approach for modeling LoS channels of H-MIMO from an EM perspective. In order to solve the problem faced by \cite{Gong2023Generalized,Gong2023HMIMO}, we propose a transmit-receive parameter separable channel modeling approach, where the original channel model, in which TX and RX parameters are all coupled together, is decomposed into two individual components that correspond to the TX and RX.  The proposed channel model requires only the knowledge of the center-points distance and direction between the TX and RX surfaces, and the local position among the antenna elements. These parameters are generally easy to obtain without the need for several repeated parameter measurements. 
Based on the proposed separable modeling approach, an easy-to-use EM-domain channel model is presented, and it is shown to be convenient for the analysis and design of H-MIMO.

The rest of the paper is organized as follows. In Section \ref{SectionSystemModel}, we present the existing EM-domain LoS channel model. In Section \ref{SectionChannelModel}, we propose the transmit-receive parameter separable EM channel modeling approach. The channel capacity is evaluated in Section \ref{SectionCCA}. Numerical results are presented in Section \ref{SectionNR}, and conclusions are drawn in Section \ref{SectionCON}.

\section{Existing EM-domain LoS Channel Model}
\label{SectionSystemModel}

We consider a point-to-point H-MIMO system that includes a TX and a RX, which are equipped with nearly continuous H-MIMO surfaces with aperture $A_{T}$ and $A_{R}$, respectively. For modeling purposes, we assume that the two surfaces consist of an infinite number of antenna elements with an infinitesimal element spacing. We assume that each antenna element is capable of generating and sensing any polarization of the transmit and receive waves. To characterize the H-MIMO system, an EM-domain system model is needed to fully capture the essence of wave propagation. To this end, we depart from the fundamental relation between the electric field $\bm{e}$ and the current distribution $\bm{j}$ at the transmission, which can be expressed as  
\begin{align}
	\label{eq:EF-CD-Relation}
	{\bm{e}}\left( {{{\bm{r}}}} \right) = \frac{\eta}{2 \lambda} \int_{A_{T}} {{\bm{G}}\left( {{{\bm{r}}},{{\bm{s}}}} \right)} {\bm{j}}\left( {{{\bm{s}}}} \right)  {\rm{d}}{{\bm{s}}},
\end{align}
where $\lambda$ and $\eta$ are the free-space wavelength and impedance, respectively; $\bm{s} \in \mathbb{R}^{3 \times 1}$ and $\bm{r} \in \mathbb{R}^{3 \times 1}$ denote the point positions on the TX and RX surfaces, respectively; ${{\bm{G}}\left( {{{\bm{r}}},{{\bm{s}}}} \right)} \in \mathbb{C}^{3 \times 3}$ is the tensor Green function, given by
\begin{align}
	\nonumber
	&{{\bm{G}}\left( {{{\bm{r}}},{{\bm{s}}}} \right)} = {\bm{G}}\left( \bm{d} \right)
	= \frac{-i}{{4\pi d}} \left[ \left( {1 + \frac{i}{{{k_0}d}} - \frac{1}{{k_0^2{d^2}}}} \right) {{\bm{I}}_3} \right. \\
	\label{eq:TGF1}
	& \qquad \qquad \qquad \qquad \left. + \left( {\frac{3}{{k_0^2{d^2}}} - \frac{{3i}}{{{k_0}d}} - 1} \right) \frac{ \bm{d} \bm{d}^{H} } {{{d^2}}} \right] 
	e^{i{k_0} d}, 
\end{align}
where $i^2 = -1$, $\bm{I}_3$ is the $3 \times 3$ identity matrix; $k_{0} = 2 \pi / \lambda$ is the wavenumber; $\bm{d} = \bm{r} - \bm{s}$ and $d = \left\| \bm{d} \right\|_{2}$ with $\| \cdot \|_{2}$ denoting the $l_2$ norm of a vector. 
It is worth mentioning that the tensor Green function privides the channel response of an LoS channel between two point positions, which offers accurate channel response coefficients for any distance range, spanning from the near field to the middle and far fields. In the far field case, $d \gg \max\{A_{T}, A_{R}\} / \lambda$, ${\bm{G}}\left( \bm{d} \right)$ can be simplified to ${\bm{G}}\left( \bm{d} \right) = \frac{-i e^{i{k_0} d}}{{4\pi d}} \left( {{\bm{I}}_3} + \frac{ \bm{d} \bm{d}^{H} } {{{d^2}}} \right)$.

For modeling purposes, the TX and RX surfaces are assumed to have $N$ and $M$ discrete antenna elements whose surface areas are $a_{T}$ and $a_{R}$, respectively. Thus, the communication model that expresses the measured electric field as a function of the surface current distribution (assumed uniformly distributed over each antenna element area) is given by \cite{Gong2023HMIMO}
\begin{align}
	\label{eq:CommModel}
	\bm{e}_{m} = \bm{H}_{mn} \bm{j}_{n} + a_{R} \bm{w}_{m},
\end{align}
where $\bm{w}_{m} \sim \mathcal{CN}(0, \sigma_{w}^{2} {{\bm{I}}_3})$ denotes the noise vector and the channel matrix $\bm{H}_{mn}$ is given by \cite{Gong2023Generalized,Gong2023HMIMO}
\begin{align}
	\nonumber
	\bm{H}_{mn} \triangleq \frac{\eta}{2 \lambda} \int\nolimits_{a_R} \int\nolimits_{a_T} {{\bm{G}}\left( {{{\bm{r}}_m},{{\bm{s}}_n}} \right)} {\rm{d}}{{\bm{s}}_n} {\rm{d}}{{\bm{r}}_m} 
	\approx \frac{\eta}{2 \lambda}  a_R a_T \bm{G} \left( \bm{d}_{mn} \right), 
\end{align}
where the approximation holds for infinitesimally small antenna elements; $\bm{d}_{mn}$ denotes the distance between the two centers of the $n$th transmit and $m$th receive surface elements, ${\bm{G}}\left( \bm{d}_{mn} \right)$ is obtained by replacing $\bm{d}$ and $d$ in \eqref{eq:TGF1} with $\bm{d}_{mn}$ and $d_{mn} = \|\bm{d}_{mn}\|_2$, respectively.
Collecting the channel of the $mn$-pair in a channel matrix, the point-to-point H-MIMO channel can be approximated as \cite{Gong2023HMIMO}
\begin{align}
	\label{eq:ChannelModel-HMIMO}
	\bm{H} \approx \frac{\eta}{2 \lambda}  a_R a_T \bm{G},
\end{align}
where $\bm{G}$ is an $3M \times 3N$ matrix whose elements are $\bm{G} \left( \bm{d}_{mn} \right)$.

\textit{Remark 1:}
It is worth noting that the channel matrix in \eqref{eq:ChannelModel-HMIMO} is capable of providing a feasible channel model for H-MIMO systems that is mainly useful for numerical calculations. It, however, fails to shed fundamental engineering insights. Specifically, the channel model relies on the distance $d_{mn}$ and the direction $\bm{d}_{mn}/d_{mn}$ between each $mn$-pair. Once the distances/directions are known, the channel matrix can be computed. 
However, in most cases, the acquisition of all distances/directions may not be straightforward. Two options for obtaining the distances/directions are: 1) Computing $d_{mn}$ and $\bm{d}_{mn}$ from the global positions $\bm{r}_{m}$ and $\bm{s}_{n}$ of all antenna elements, i.e., $\bm{d}_{mn} = \bm{r}_{m} - \bm{s}_{n}$ and $d_{mn} = \|\bm{d}_{mn}\|_{2}$. 2) Directly estimating $d_{mn}$ and $\bm{d}_{mn}$ through measurements of the distances and angles of arrival. 
Both approaches are inefficient in practical applications due to the high overhead that is needed to estimate all these parameters. Particularly, at least $M+N$ measurements are needed, which entails a high complexity, especially for H-MIMO systems.

To deal with this problem, we propose to model the channel via a transmit-receive parameter separation approach that is concise but accurate in capturing the H-MIMO channels.

\section{Proposed Transmit-Receive Parameter Separable EM-Domain Channel Modeling}
\label{SectionChannelModel}

\begin{figure}[t!]
	\centering
	\includegraphics[height=4.6cm, width=6cm]{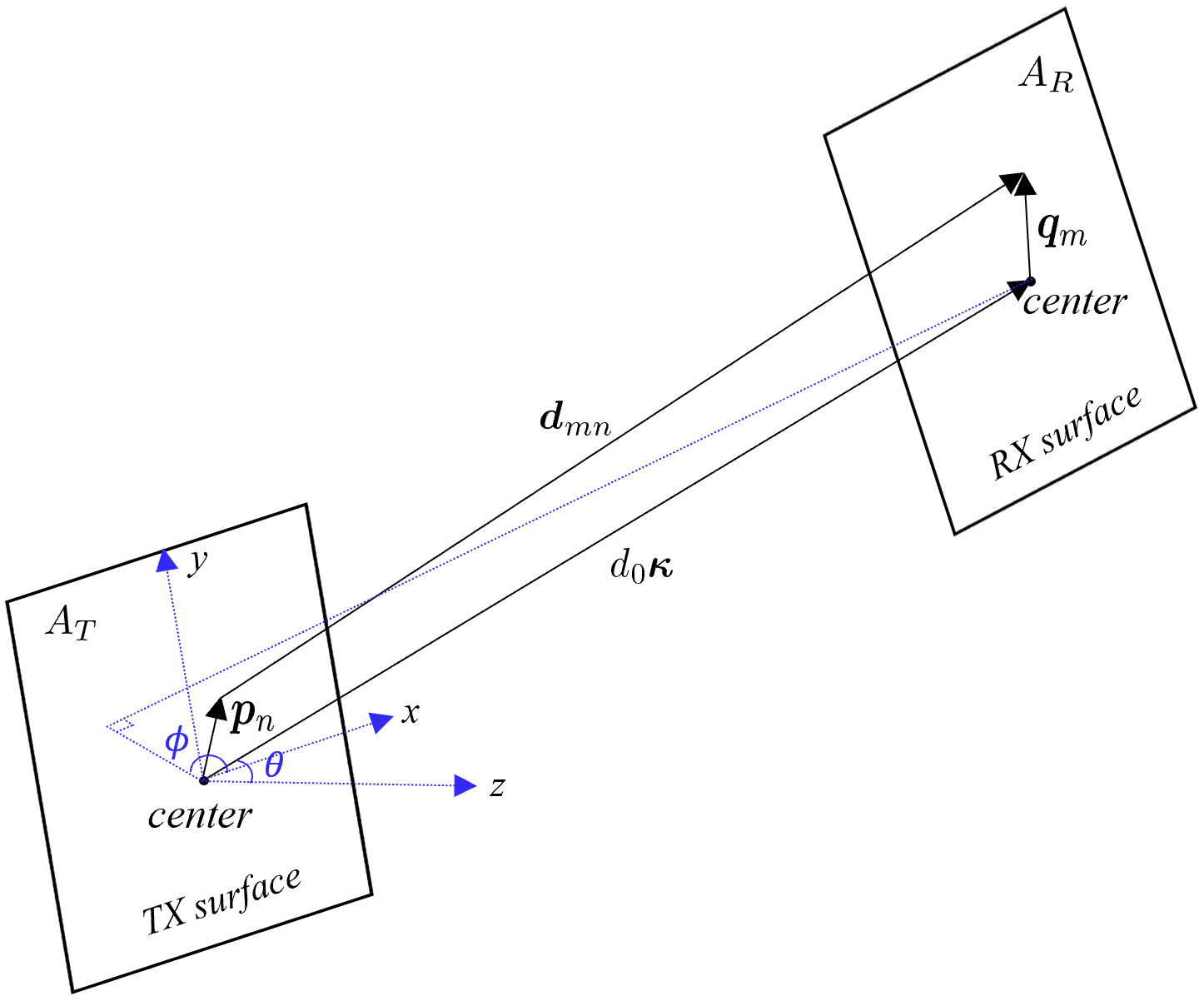}
	\caption{Separable modeling approach for a H-MIMO system.}
	\label{fig:ElementVector}
\end{figure}

\subsection{Partially Separable EM-Domain Channel Modeling}
For achieving the separability of the parameters, we propose to decouple $d_{mn}$ and $\bm{d}_{mn}$ using a tractable representation. To this end, we select two origins at the TX and RX surfaces, and denote the distance between them by $d_{0}$. Also, we introduce the wavevector $\bm{\kappa} = [\sin \theta \cos \phi \;\;\sin \theta \sin \phi \;\;\cos \theta]^{T}$ of the origin-to-origin direction, where $\theta$ and $\phi$ are the elevation angle and the azimuth angle, respectively, evaluated with respect to the TX surface. In addition, we define the local position vector $\bm{p}_{n} \in A_{T}$ and the local position vector $\bm{q}_{m} \in A_{R}$ to indicate the $n$-th and $m$-th antenna element of the TX surface and RX surface compared to their origins, respectively. It is emphasized that the distance vector of each $mn$-pair can be represented through $d_0$, $\bm{\kappa}$, $\bm{p}_{n}$ and $\bm{q}_{m}$, as demonstrated in Fig. \ref{fig:ElementVector}. 
Specifically, $\bm{d}_{mn}$ can thus be expressed as 
\begin{align}
	\label{eq:dmn}
	\bm{d}_{mn} = d_{0} \bm{\kappa} - \bm{p}_{n} + \bm{q}_{m}.
\end{align}
The expression in \eqref{eq:dmn} decouples the distance (and direction) of each $mn$-pair into three components. Compared with the direct measurement of $\bm{d}_{mn}$ that needs to be performed for each $mn$-pair, $\bm{p}_{n}$ and $\bm{q}_{m}$ are local positions that correspond to the TX and RX surfaces, respectively, which are easy to obtain by knowing the indices of the antenna elements and the corresponding element spacing without performing a large number of measurements. Only the acquisition of $d_{0} \bm{\kappa}$ requires some channel measurements. 
The proposed representation facilitates the establishment of an easy-to-use channel model. 

Based on the representation in \eqref{eq:dmn}, we obtain
\begin{align}
	\nonumber
	\bm{d}_{mn} \bm{d}_{mn}^{H} 
	&= d_0^2{\bm{\kappa }}{{\bm{\kappa }}^H} 
	+ {d_0}{\bm{\kappa }}{\left( {{{\bm{q}}_m} - {{\bm{p}}_n}} \right)^H} + {d_0}\left( {{{\bm{q}}_m} - {{\bm{p}}_n}} \right){{\bm{\kappa }}^H} \\
	\label{eq:dmndmn}
	&+ \left( {{{\bm{q}}_m} - {{\bm{p}}_n}} \right){\left( {{{\bm{q}}_m} - {{\bm{p}}_n}} \right)^H},
\end{align}
and the $mn$-pair distance can be expressed as
\begin{align}
	\nonumber
	&d_{mn}^2 = {\rm{trace}} \left( \bm{d}_{mn} \bm{d}_{mn}^{H} \right) \\
	\label{eq:dmn2}
	&= d_0^2 + 2{d_0}{\left( {{{\bm{q}}_m} - {{\bm{p}}_n}} \right)^H}{\bm{\kappa }} + {\left\| {{{\bm{q}}_m} - {{\bm{p}}_n}} \right\|^2} 
	= \alpha_{mn}^{2} d_0^2,
\end{align}
where $\alpha_{mn} = \left[ {1 + 2\frac{{{{\left( {{{\bm{q}}_m} - {{\bm{p}}_n}} \right)}^H}}}{{{d_0}}}{\bm{\kappa }} + {{\left( {\frac{{\left\| {{{\bm{q}}_m} - {{\bm{p}}_n}} \right\|}}{{{d_0}}}} \right)}^2}} \right]^{\frac{1}{2}}$.

Since $\| \bm{\kappa} \| = 1$ and applying the Cauchy-Schwarz inequality, the following lower bound for $d_{mn}^2$ can be obtained
\begin{align}
	\nonumber
	d_{mn}^2 
	&\ge d_0^2\left[ {1 + 2\frac{{{{\left( {{{\bm{q}}_m} - {{\bm{p}}_n}} \right)}^H}}}{{{d_0}}}{\bm{\kappa }} + {{\left( {\frac{{{{\left( {{{\bm{q}}_m} - {{\bm{p}}_n}} \right)}^H}}}{{{d_0}}}{\bm{\kappa }}} \right)}^2}} \right] \\
	\label{eq:lowerbound}
	&= \gamma_{mn}^{2} d_0^2, 
\end{align}
where $\gamma_{mn} = {1 + \frac{{{{\left( {{{\bm{q}}_m} - {{\bm{p}}_n}} \right)}^H}}}{{{d_0}}}{\bm{\kappa }}}$.

Substituting $\bm{d}_{mn} \bm{d}_{mn}^{H}$ and $d_{mn}$ into \eqref{eq:TGF1}, we get 
\begin{align}
	\nonumber
	&{\bm{G}}\left( \bm{d}_{mn} \right) = {\bm{G}}\left( \bm{q}_{m}, \bm{p}_{n}, \bm{\kappa}, d_{0} \right) \\
	\nonumber
	&= \frac{-i}{{4\pi \alpha_{mn} d_0}} \left( \bm{A}_{mn}^{(1)} + \bm{A}_{mn}^{(2)} + \bm{A}_{mn}^{(3)} + \bm{A}_{mn}^{(4)}  \right) e^{i{k_0} \alpha_{mn} d_0} \\
	\nonumber
	&\approx \frac{-i}{{4\pi \gamma_{mn} d_0}} \left( \bm{A}_{mn}^{(1)} + \bm{A}_{mn}^{(2)} + \bm{A}_{mn}^{(3)} + \bm{A}_{mn}^{(4)}  \right) e^{i{k_0} \gamma_{mn} d_0} \\
	\label{eq:TensorGreenFunction1}
	&\triangleq \frac{-i}{{4\pi \gamma_{mn} d_0}} \bm{A}_{mn}  e^{i{k_0} d_0}  e^{ i{k_0} \bm{q}_{m}^{H} \bm{\kappa}} 
	 e^{- i{k_0} \bm{p}_{n}^{H} \bm{\kappa}},
\end{align}
where the approximation is obtained by replacing $\alpha_{mn}$ with its lower bound $\gamma_{mn}$; $\bm{A}_{mn} \triangleq \bm{A}_{mn}^{(1)} + \bm{A}_{mn}^{(2)} + \bm{A}_{mn}^{(3)} + \bm{A}_{mn}^{(4)}$ with each component given by
\begin{align}
	\label{eq:Amn1}
	\bm{A}_{mn}^{(1)} &= \omega_{mn}^{(1)}  {{\bm{I}}_3}, \\
	\label{eq:Amn2}
	\bm{A}_{mn}^{(2)} &= \omega_{mn}^{(2)}  \frac{ \bm{\kappa} \bm{\kappa}^{H} } {{\gamma_{mn}^{2}}}, \\
	\label{eq:Amn3}
	\bm{A}_{mn}^{(3)} &= \omega_{mn}^{(2)}  \frac{ \bm{\kappa} \left(\bm{q}_{m} - \bm{p}_{n} \right)^{H} + \left(\bm{q}_{m} - \bm{p}_{n} \right) \bm{\kappa}^{H} } {{\gamma_{mn}^{2} d_0}}, \\
	\label{eq:Amn4}
	\bm{A}_{mn}^{(4)} &= \omega_{mn}^{(2)}  \frac{ \left(\bm{q}_{m} - \bm{p}_{n} \right) \left(\bm{q}_{m} - \bm{p}_{n} \right)^{H} } {{\gamma_{mn}^{2} d_0^2}},
\end{align}
and 
\begin{align}
	\label{eq:wmn1}
	\omega_{mn}^{(1)} = {1 + \frac{i}{{{k_0} \gamma_{mn} d_0}} - \frac{1}{{k_0^2 \gamma_{mn}^{2} d_0^2}}}, \\
	\label{eq:wmn2}
	\omega_{mn}^{(2)} = {\frac{3}{{k_0^2 \gamma_{mn}^{2} d_0^2}} - \frac{{3i}}{{{k_0} \gamma_{mn} d_0}} - 1}.
\end{align}

By direct inspection of \eqref{eq:TensorGreenFunction1}, we see that the coupled phase term $e^{i{k_0} \alpha_{mn} d_0}$ is separated into two individual phase terms $e^{- i{k_0} \bm{q}_{m}^{H} \bm{\kappa}}$ and $e^{- i{k_0} \bm{p}_{n}^{H} \bm{\kappa}}$, which facilitates the channel modeling of H-MIMO. However, it is difficult to obtain a completely separable expression for $\bm{A}_{mn}$, e.g., writing $\bm{A}_{mn} = \bm{A}_{mn}(\bm{q}_{m}) + \bm{A}_{mn}(\bm{p}_{n})$ with $\bm{A}_{mn}(\bm{q}_{m})$ and $\bm{A}_{mn}(\bm{p}_{n})$ being the components expressed only in terms of $\bm{q}_{m}$ and $\bm{p}_{n}$, respectively, due to the intractable expression for ${{1/\gamma_{mn}}}$ and $\bm{A}_{mn}^{(4)}$. 
However, the proposed approach is still superior compared with the original fully coupled tensor Green function in \eqref{eq:TGF1}, which is shown to facilitate the development of a H-MIMO channel model, as detailed next.

\begin{figure*}
	\begin{align}
		\label{eq:TGF3}
		\bm{G} = \frac{-i {e^{i{k_0}{d_0}}}}{{4 \pi d_0}}  \left[ {\begin{array}{*{20}{c}}
				{\frac{{{\bm{A}}_{11}}}{{\gamma_{11}}}  {e^{i{k_0}{\bm{q}}_1^H{\bm{\kappa }}}}  {e^{ - i{k_0}{\bm{p}}_1^H{\bm{\kappa }}}}}&{\frac{{{\bm{A}}_{12}}}{{\gamma_{12}}}  {e^{i{k_0}{\bm{q}}_1^H{\bm{\kappa }}}}  {e^{ - i{k_0}{\bm{p}}_2^H{\bm{\kappa }}}}}& \cdots &{\frac{{{\bm{A}}_{1N}}}{{\gamma_{1N}}}  {e^{i{k_0}{\bm{q}}_1^H{\bm{\kappa }}}}  {e^{ - i{k_0}{\bm{p}}_N^H{\bm{\kappa }}}}}\\
				{\frac{{{\bm{A}}_{21}}}{{\gamma_{21}}}  {e^{i{k_0}{\bm{q}}_2^H{\bm{\kappa }}}}  {e^{ - i{k_0}{\bm{p}}_1^H{\bm{\kappa }}}}}&{\frac{{{\bm{A}}_{22}}}{{\gamma_{22}}}  {e^{i{k_0}{\bm{q}}_2^H{\bm{\kappa }}}}  {e^{ - i{k_0}{\bm{p}}_2^H{\bm{\kappa }}}}}& \cdots &{\frac{{{\bm{A}}_{2N}}}{{\gamma_{2N}}}  {e^{i{k_0}{\bm{q}}_2^H{\bm{\kappa }}}}  {e^{ - i{k_0}{\bm{p}}_N^H{\bm{\kappa }}}}}\\
				\vdots & \vdots & \ddots & \vdots \\
				{\frac{{{\bm{A}}_{M1}}}{{\gamma_{M1}}}  {e^{i{k_0}{\bm{q}}_M^H{\bm{\kappa }}}}  {e^{ - i{k_0}{\bm{p}}_1^H{\bm{\kappa }}}}}&{\frac{{{\bm{A}}_{M2}}}{{\gamma_{M2}}}  {e^{i{k_0}{\bm{q}}_M^H{\bm{\kappa }}}}  {e^{ - i{k_0}{\bm{p}}_2^H{\bm{\kappa }}}}}& \cdots &{\frac{{{\bm{A}}_{MN}}}{{\gamma_{MN}}}  {e^{i{k_0}{\bm{q}}_M^H{\bm{\kappa }}}}  {e^{ - i{k_0}{\bm{p}}_N^H{\bm{\kappa }}}}}
		\end{array}} \right],
	\end{align}
{\noindent} \rule[0pt]{18cm}{0.05em}
\end{figure*}

Specifically, using \eqref{eq:TensorGreenFunction1}, the H-MIMO channel matrix in \eqref{eq:TGF3} can be obtained. Furthermore, we obtain the following decoupled expression
\begin{align}
	\label{eq:TensorGreenMatrix}
	\bm{G} 
	= \frac{-i{e^{i{k_0}{d_0}}}}{{4 \pi d_0}} \left( {{{\bm{\theta }}_R}{\bm{\theta }}_T^H \otimes {{\bm{\varepsilon }}_3}{\bm{\varepsilon }}_3^H} \right) \odot {\bm{A}},
\end{align}
where $\otimes$ and $\odot$ are the Kronecker product and the Hadamard product, respectively; $\bm{\theta}_{T}$ and $\bm{\theta}_{R}$ are the TX and RX array response vectors, expressed as
\begin{align}
	\label{eq:TArrayVector}
	\bm{\theta}_{T} &= \left[{e^{ i{k_0}{\bm{p}}_1^H{\bm{\kappa }}}} \;\; {e^{ i{k_0}{\bm{p}}_2^H{\bm{\kappa }}}} \;\; \cdots \;\; {e^{ i{k_0}{\bm{p}}_N^H{\bm{\kappa }}}} \right]^{T}, \\
	\label{eq:RArrayVector}
	\bm{\theta}_{R} &= \left[{e^{  i{k_0}{\bm{q}}_1^H{\bm{\kappa }}}} \;\; {e^{  i{k_0}{\bm{q}}_2^H{\bm{\kappa }}}} \;\; \cdots \;\; {e^{  i{k_0}{\bm{q}}_M^H{\bm{\kappa }}}} \right]^{T},
\end{align}
$\bm{\varepsilon}_{3} = \left[ 1 \;\; 1 \;\; 1 \right]^{T}$ and $\bm{A}$ is defined as
\begin{align}
	\label{eq:Amatrix}
	{\bm{A}} = \left[ {\begin{array}{*{20}{c}}
			\frac{{{\bm{A}}_{11}}}{{\gamma_{11}}} & \frac{{{\bm{A}}_{12}}}{{\gamma_{12}}} & \cdots & \frac{{{\bm{A}}_{1N}}}{{\gamma_{1N}}} \\
			\frac{{{\bm{A}}_{21}}}{{\gamma_{21}}} & \frac{{{\bm{A}}_{22}}}{{\gamma_{22}}} & \cdots & \frac{{{\bm{A}}_{2N}}}{{\gamma_{2N}}} \\
			\vdots & \vdots & \ddots & \vdots \\
			\frac{{{\bm{A}}_{M1}}}{{\gamma_{M1}}} & \frac{{{\bm{A}}_{M2}}}{{\gamma_{M2}}} & \cdots & \frac{{{\bm{A}}_{MN}}}{{\gamma_{MN}}} 
	\end{array}} \right], 
\end{align}
which can be expressed as the sum of four components, i.e., $\bm{A} = \bm{A}_{1} + \bm{A}_{2} + \bm{A}_{3} + \bm{A}_{4}$, with $\bm{A}_{i}$, $i \in \{1, 2, 3, 4\}$ corresponding to $\bm{A}_{mn}^{(i)}$ in \eqref{eq:Amn1}, \eqref{eq:Amn2}, \eqref{eq:Amn3} and \eqref{eq:Amn4}, respectively.

\textit{Remark 2:}
Compared with the representation based on the tensor Green function, the proposed H-MIMO channel model obtained from \eqref{eq:TensorGreenMatrix} has two main advantages. First, the estimation of the H-MIMO channel using \eqref{eq:TensorGreenMatrix} requires far less than $M+N$ parameter measurements. Specifically, $\bm{G}$ in \eqref{eq:TensorGreenMatrix} is determined by $\bm{p}_{n}$, $n = 1, 2, \cdots, N$, $\bm{q}_{m}$, $m = 1, 2, \cdots, M$, and $d_{0} \bm{\kappa}$. The acquisition of $d_{0} \bm{\kappa}$ needs one measurement. Furthermore, $\bm{p}_{n}$ and $\bm{q}_{m}$ can be directly calculated given the indices of the antenna elements and the inter-distance between the antenna elements, without the need of channel measurements, which greatly facilitates the application of the proposed channel model.
Second, \eqref{eq:TensorGreenMatrix} showcases that the H-MIMO channel can be formulated in term of the TX and RX array response vectors, which facilitates analysis and signal processing developments. 
It is worth mentioning that the these advantages are obtained by the proposed channel model at the expenses of slightly sacrificing the accuracy in the near field (as shown in the numerical evaluations). 


\begin{table*}[!t]
	\footnotesize
	\renewcommand{\arraystretch}{1.2}
	\caption{Comparison between the proposed channel model and the conventional channel model.}
	\label{table}
	\centering
	\begin{tabular}{c|c|c|c}
		\hline
		Channel model & Tensor Green function \eqref{eq:TGF1} \cite{Gong2023Generalized, Gong2023HMIMO} & \multicolumn{2}{c}{Our separated model \eqref{eq:TensorGreenMatrix} and \eqref{eq:TensorGreenMatrix1}} \\
		\hline
		Parameters needed & $\bm{d}_{mn}$, $\forall m, n$ & $\bm{q}_{m}$, $\forall m$, \; $\bm{p}_{n}$, $\forall n$ & $d_{0} \bm{\kappa}$ \\
		\hline 
		Acquisition method & distance/direction measurements & direct numerical calculations & distance/direction measurements \\
		\hline
		Number of measurements & $\ge M+N$ & $/$ & $1$ \\
		\hline
	\end{tabular}
\end{table*}

\subsection{Fully Separable EM-Domain Channel Modeling}

In this section, we show that the proposed channel model can be further simplified to a fully separable model when $d_{0}$ is large, namely, in the far field case. Specifically, we assume $d_{0} \gg \max\{A_{T}, A_{R}\} / \lambda$. For a fixed $\lambda$, we obtain 
\begin{align}
	\label{eq:Condition1}
	d_{0} \gg \left\| \bm{q}_{m} - \bm{p}_{n} \right\|^{2} \ge \left( \bm{q}_{m} - \bm{p}_{n} \right)^{H} \bm{\kappa},
\end{align}
such that $\alpha_{mn}$ and $\gamma_{mn}$ simplify to
\begin{align}
	\label{eq:Condition2}
	\alpha_{mn} \approx \gamma_{mn} \approx 1.
\end{align}

By using \eqref{eq:Condition1} and \eqref{eq:Condition2}, we also obtain
\begin{align}
	\nonumber
	\omega_{mn}^{(1)} &\approx 1, 
	\;\;\;
	\omega_{mn}^{(2)} \approx - 1, \\
	\nonumber
	\bm{A}_{mn}^{(1)} &\approx  {{\bm{I}}_3}, 
	\;
	\bm{A}_{mn}^{(2)} \approx - \bm{\kappa} \bm{\kappa}^{H}, \;
	\bm{A}_{mn}^{(3)} \approx 
	\bm{A}_{mn}^{(4)} \approx \bm{0}.
\end{align}

Therefore, we evince that $\bm{A}_{mn}^{(3)}$ and $\bm{A}_{mn}^{(4)}$, which depend on $\bm{q}_{m}$ and $\bm{p}_{n}$, tend to be zero, implying that
\begin{align}
	\label{eq:AmnmatrixFF}
	\bm{A}_{mn} &\approx \bm{A}_{mn}^{(1)} + \bm{A}_{mn}^{(2)}  
	= {{{\bm{I}}_3} - {\bm{\kappa }}{{\bm{\kappa }}^{H}}}
	\triangleq \bm{C},
\end{align}
which is only determined by $d_{0}$ and $\bm{\kappa}$, while being invariant with respect to the positions of the TX and RX antenna elements. Therefore, \eqref{eq:Amatrix} simplifies to 
\begin{align}
	\label{eq:Amatrix1}
	\bm{A} = \bm{\varepsilon}_{M} \bm{\varepsilon}_{N}^{H} \otimes \bm{C},
\end{align}
where $\bm{\varepsilon}_{M}$ and $\bm{\varepsilon}_{N}$ are $M \times 1$ and $N \times 1$ all-one vectors, respectively. Substituting \eqref{eq:Amatrix1} into \eqref{eq:TensorGreenMatrix}, and exploiting the equality $\left( \bm{X} \otimes \bm{Y} \right) \odot \left( \bm{Z} \otimes \bm{W} \right) = \left( \bm{X} \odot \bm{Z} \right) \otimes \left( \bm{Y} \odot \bm{W} \right)$, we obtain the following far field approximation for $\bm{G}$ 
\begin{align}
	\nonumber
	\bm{G}_{{\rm{FF}}} 
	&= \frac{-i {e^{i{k_0}{d_0}}}}{{4 \pi d_0}} \left( {{{\bm{\theta }}_R}{\bm{\theta }}_T^H \otimes {{\bm{\varepsilon }}_3}{\bm{\varepsilon }}_3^H} \right) \odot \left( \bm{\varepsilon}_{M} \bm{\varepsilon}_{N}^{H} \otimes \bm{C} \right) \\
	\label{eq:TensorGreenMatrix1}
	&= \frac{-i {e^{i{k_0}{d_0}}}}{{4 \pi d_0}} \bm{\theta}_{R} \bm{\theta}_{T}^{H} \otimes {\left( {{{\bm{I}}_3} - {\bm{\kappa }}{{\bm{\kappa }}^{H}}} \right)}.
\end{align}

It is worth noting that ${\mathrm{rank}} \left( {{\bm{I}}_3} - \bm{\kappa} \bm{\kappa}^{T} \right) = 2$, indicating that the direction of oscillation of the EM wave is constrained to be perpendicular to its propagation direction. 
It is noteworthy that the derived far field channel model in \eqref{eq:TensorGreenMatrix1} is a two-polarization model. The widely studied one-polarization model in conventional LoS massive MIMO is given by $\beta \bm{\theta}_{R} \bm{\theta}_{T}^{H}$ (where $\beta$ accounts for the path loss), which is a special case of \eqref{eq:TensorGreenMatrix1}.

The explicit comparison between the proposed channel model and the conventional tensor Green function channel model is given in Table \ref{table}.

\addtolength{\topmargin}{0.032in}


\section{Channel Capacity Evaluation}
\label{SectionCCA}

Based on \eqref{eq:CommModel}, the H-MIMO communication model can be formulated as 
\begin{align}
	\bm{e} = \bm{H} \bm{j} + a_{R} \bm{w},
\end{align}
where $\bm{e}$, $\bm{j}$ and $\bm{w}$ are obtained by collecting $\bm{e}_{m}$, $\bm{j}_{n}$ and $\bm{w}_{m}$, respectively, and the channel matrix $\bm{H}$ is given by \eqref{eq:ChannelModel-HMIMO}. Let us assume that the current distribution $\bm{j}$ is expressed as a weighted sum of transmit patterns $\bm{u}_{p} \in \mathbb{C}^{N \times 1}$, and that the electric field is expressed as a weighted sum of receive patterns $\bm{v}_{p} \in \mathbb{C}^{M \times 1}$, where the weights correspond to the transmit symbols $c_{p}$ and receive symbols $\hat{c}_{p}$, respectively. Denoting $\bm{T} = [\bm{u}_{1}, \bm{u}_{2}, \cdots, \bm{u}_{P}]$, $\bm{R} = [\bm{v}_{1}, \bm{v}_{2}, \cdots, \bm{v}_{P}]$, $\bm{c} = [c_{1}, c_{2}, \cdots, c_{P}]^{T}$ and $\hat{\bm{c}} = [\hat{c}_{1}, \hat{c}_{2}, \cdots, \hat{c}_{P}]^{T}$, we get the communication model between the TX and RX symbols \cite{Gong2023HMIMO}
\begin{align}
	\hat{\bm{c}} = \frac{\eta}{2 \lambda}  a_R a_T \bm{R}^{H} \bm{G} \bm{T} \bm{c} + a_R \bm{R}^{H} \bm{w}.
\end{align}

Using the bilinear decomposition of $\bm{G} = \bm{R} \bm{D} \bm{T}^{H}$, where $\bm{D}$ is a diagonal matrix with its $p$-th element given by $\gamma_{p} \approx a_{R} a_{T} \bm{v}_{p}^{H} {{\bm{G}}} \bm{u}_{p}$ \cite{Gong2023HMIMO}, the channel capacity can be written as
\begin{align}
	\label{eq:Capacity}
	C = \sum\limits_{p = 1}^P {\log_{2} \left( {1 + \mu  \mathrm{SNR}  \gamma_{p}^2} \right)} 
\end{align}
by considering a simple uniform power allocation scheme for all the independent (parallel) transmission channels, where $\mu = \frac{\eta^2}{4 \lambda^2}$ and $\mathrm{SNR} = \frac{\mathcal{P}_{t}}{P a_{R} \sigma_w^2}$ denotes the average transmit signal-to-noise ratio (SNR) with $\mathcal{P}_{t}$ being the total transmit power. 
The parameter $\gamma_p$ can be obtained by applying the singular value decomposition to $\bm{G} = \bm{U} \bm{\Sigma} \bm{V}^{H}$ if $\bm{T}$ and $\bm{R}$ are not available. 
By setting $\bm{R} = \frac{\bm{U}(:, 1:P)}{\sqrt{a_{R}}}$, $\bm{T} = \frac{\bm{V}(:, 1:P)}{\sqrt{a_{T}}}$, $\gamma_{p}$ can be determined according to $\bm{D} = \sqrt{a_{R} a_{T}} \bm{\Sigma}(1:P, 1:P)$. 

\section{Numerical Analysis}
\label{SectionNR}

Here, we perform a numerical study to evaluate the accuracy of the proposed H-MIMO channel model in terms of normalized mean-squared error (NMSE), which is expressed as ${\rm{NMSE}} = \| \hat{\bm{H}} - \bm{H} \|_{F}^{2} / \| \bm{H} \|_{F}^{2}$, 
where $\bm{H}$ and $\hat{\bm{H}}$ are the original channel matrix in \eqref{eq:ChannelModel-HMIMO} and the proposed channel matrix using the separable channel model, respectively. 
In addition, we evaluate the proposed channel model with respect to the channel capacity $C$ in \eqref{eq:Capacity}, and compare it with the channel capacity of the original channel model. 
Specifically, the NMSE and the channel capacity is evaluated in terms of the TX-RX distance and the number of TX and RX antenna elements. The operating frequency is 2.4 GHz, and the TX and RX surfaces are parallel to each other with their center-points perfectly aligned.
Specifically, we consider the following setups:
\begin{itemize}
	\item 
	``Original channel model (OCM)", where $\bm{H}$ is obtained by \eqref{eq:ChannelModel-HMIMO} and $\bm{G}$ is obtained by \eqref{eq:TGF1};
	
	\item 
	``Partially separable channel model (PSCM)", where $\hat{\bm{H}}$ is obtained by \eqref{eq:ChannelModel-HMIMO} and $\bm{G}$ is obtained by \eqref{eq:TensorGreenMatrix} with $\bm{A} = \bm{A}_{1} + \bm{A}_{2} + \bm{A}_{3} + \bm{A}_{4}$. Moreover, 
	\begin{itemize}
		\item 
		``PSCM$_{(12)}$" refers to the case $\bm{A} = \bm{A}_{1} + \bm{A}_{2}$;
		
		\item 
		``PSCM$_{(123)}$" refers to the case $\bm{A} = \bm{A}_{1} + \bm{A}_{2} + \bm{A}_{3}$;
	\end{itemize}
	
	\item 
	``Fully separable channel model (FSCM)", where $\hat{\bm{H}}$ is obtained by \eqref{eq:ChannelModel-HMIMO} and $\bm{G}$ is obtained by \eqref{eq:TensorGreenMatrix1}.
\end{itemize}

In addition, we perform the numerical study by considering the near field and the far field, which are identified by the Rayleigh distance $d_{R} = \frac{2 (D_{TX} + D_{RX})^2}{\lambda}$, where $D_{TX} = \sqrt{ \ell_{TX, h}^{2} + \ell_{TX, v}^{2} }$ with $\ell_{TX, h}$ and $\ell_{TX, v}$ are the horizontal and vertical aperture lengths of the TX; and $D_{RX}$ can be obtained in a similar manner. 

\begin{figure}[!tbp]
	\centering
	\subfloat[]{\label{fig:nMSE_TRDistance}\includegraphics[width=0.476\columnwidth]{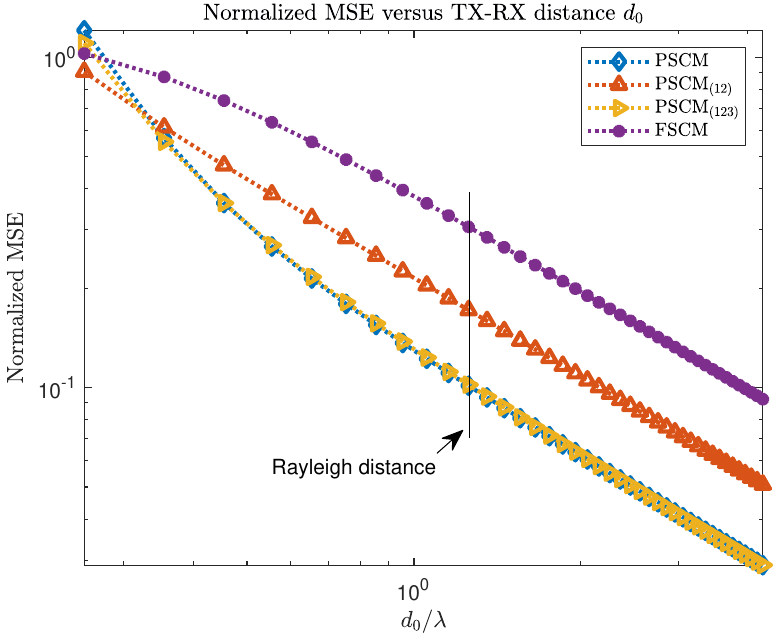}} \quad
	\subfloat[]{\label{fig:Capacity_Distance}\includegraphics[width=0.476\columnwidth]{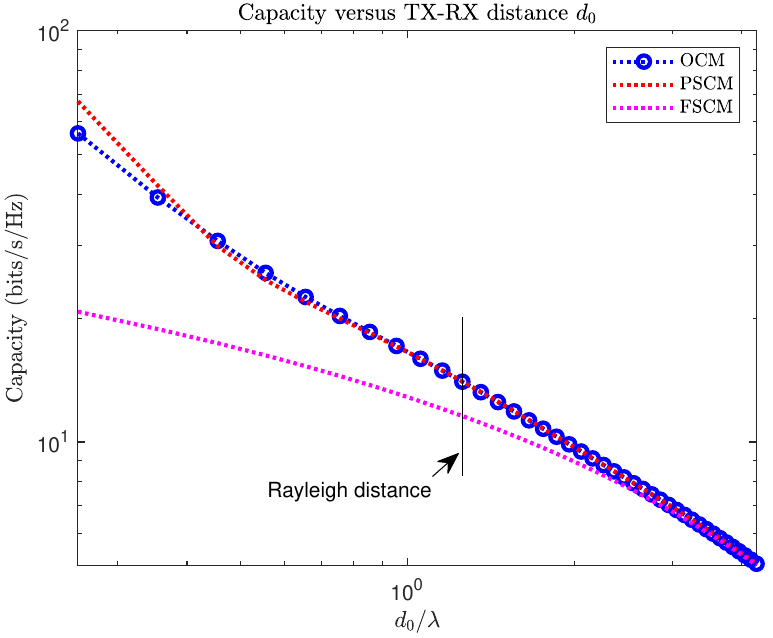}} \\
	\vspace{-0.3em}
	\caption{Evaluation of the proposed channel model versus the TX-RX distance: (a) NMSE, and (b) capacity.}
	\label{fig:nMSE_Capacity}
	\vspace{-0.4em}
\end{figure}

We first present the NMSE with respect to the TX-RX distance in Fig. \ref{fig:nMSE_TRDistance}. The number of TX and RX elements are $41 \times 41$ and $15 \times 15$, respectively. The antenna element spacing is $0.01 \lambda$, and the TX-RX distance ranges from $0.25 \lambda$ to $4.25 \lambda$, where $d_{R} = 1.25 \lambda$ identifies the boundary of the near field and far field regions. 
Three observations emerge by the inspection of Fig. \ref{fig:nMSE_TRDistance}: 1) The proposed PSCM can accurately approximate the OCM with a small NMSE, especially when $d_{0} \ge 0.35 \lambda$, 
showing the effectiveness of the proposed channel model in the near field ($0.35 \lambda \le d_{0} \le d_{R}$) and the far field ($d_{0} > d_{R}$); 2) PSCM$_{(123)}$ is a better approximation for PSCM compared with PSCM$_{(12)}$; 3) as $d_{0} \gg d_{R}$, the proposed FSCM approximated model approaches the PSCM.

In Fig. \ref{fig:Capacity_Distance}, we evaluate the channel capacity as a function of the distance $d_{0}$, by retaining the same parameter configurations. 
We find that the capacity of the proposed PSCM accurately agrees with the capacity of the OCM over the considered distance range, proving the effectiveness of the proposed approach in near field and the field conditions. Moreover, we observe that the PSCM reduces to the proposed FSCM in the far field case as $d_{0} \gg d_{R}$. 
We can additionally see that the FSCM well approaches the OCM as $d_{0} \gg d_{R}$, while it is less accurate for distances closer to the Rayleigh distance. This reveals that the conventional one-polarization LoS channel model applied in massive MIMO, a special case of the FSCM, suffers a similar lack of accuracy.

\begin{figure}[!tbp]
	\centering
	\subfloat[]{\label{fig:nMSE_ElementNum}\includegraphics[width=0.476\columnwidth]{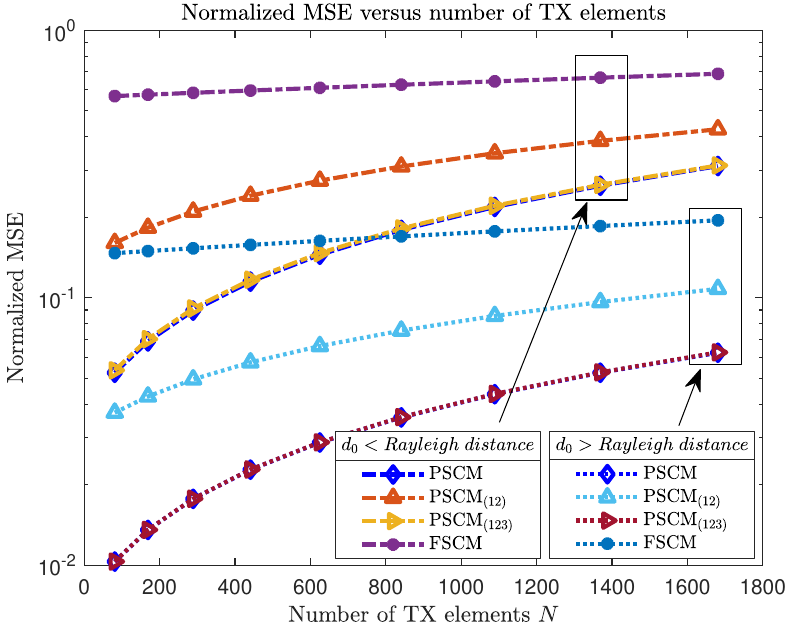}} \quad
	\subfloat[]{\label{fig:Capacity_ElementNum}\includegraphics[width=0.476\columnwidth]{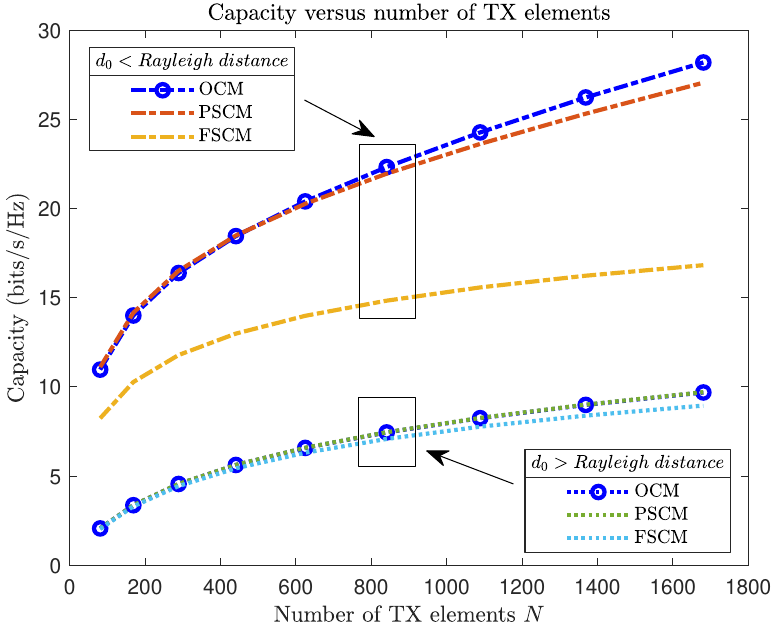}}\\
	\vspace{-0.3em}
	\caption{Evaluation of the proposed channel model versus the number of TX elements: (a) NMSE, and (b) capacity.}
	\label{fig:Capacity_nMSE}
	\vspace{-0.8em}
\end{figure}

We further evaluate the NMSE with respect to the number of TX elements $N$ in Fig. \ref{fig:nMSE_ElementNum}. $N$ varies from $81$ ($9 \times 9$) to $1681$ ($41 \times 41$). We fix the number of RX elements $M$ to $225$ ($15 \times 15$), and test two different TX-RX distances, $d_{0} < d_{R}$ and $d_{0} > d_{R}$, for analyzing the near field and the far field, respectively. 
We see from Fig. \ref{fig:nMSE_ElementNum} that: 1) PSCM$_{(123)}$ approaches PSCM in the near field and the far field, and both outperform the PSCM$_{(12)}$ and FSCM with a small NMSE; 
2) the NMSE of all models in the far field ($d_{0} > d_{R}$) is lower than that in the near field ($d_{0} < d_{R}$); 
3) as $N$ increases, the NMSE of PSCM, PSCM$_{(123)}$ and PSCM$_{(12)}$ tend to increase. As increasing $N$ results in a larger aperture area, such that $d_{0}$ tends to be shifted to the near field, this sightly degrades the channel model accuracy in terms of NMSE. 

Applying the same configurations, we present the channel capacity as a function of $N$ in Fig. \ref{fig:Capacity_ElementNum}. The proposed PSCM is capable of achieving the capacity of OCM in the near field and the far field. In addition, when $d_{0} > d_{R}$, the capacity of the FSCM tends to the capacity of the PSCM and OCM. As expected, the channel capacity increases as the transmission distance $d_{0}$ becomes shorter, and as the number of TX elements $N$ becomes larger.

In summary, the proposed PSCM provides an accurate approximation for the OCM in terms of NMSE and channel capacity over a wide range of TX-RX distances and antenna elements, while reducing the number of channel parameters to be estimated. Moreover, PSCM$_{(123)}$ provides a good approximation for PSCM with a small performance loss. Finally, the proposed PSCM reduces to the FSCM in the far field, which confirms the consistency of the proposed approach.

\section{Conclusions}
\label{SectionCON}
In this article, we studied the EM-domain LoS channel modeling for H-MIMO systems. We proposed a separable channel modeling approach that decomposes the distances/directions into individual TX and RX surface related components. This approach was shown to facilitate the development of an EM-domain LoS channel model, which can be expressed in terms of the distance between the center-points of the TX and RX surfaces, and the local TX and RX element positions. This was shown to reduce the number of parameters to estimate. We presented a capacity evaluation framework, which was used to analyze the accuracy of the proposed channel model. The numerical results demonstrated that the proposed channel model accurately approximates the original channel model and offers accurate predictions of the channel capacity.


\balance 
\bibliographystyle{IEEEtran}
\bibliography{IEEEabrv,references} 

\end{document}